\documentclass[11pt]{article}
\usepackage{fullpage}
\usepackage{cite, hyperref}
\usepackage[utf8]{inputenc}
\usepackage{amsmath}
\usepackage{siunitx, graphicx}
\usepackage{booktabs}

\newcommand{\sL}{\mathcal{L}}
\newcommand{\sB}{\mathcal{B}}

\newcommand{\lplm}{\ell^+ \ell^-}
\newcommand{\mll}{m_{\lplm}}
\newcommand{\mfl}{m_{4\ell}}

\newcommand{\PW}{\Gamma_{\Ztofl}}
\newcommand{\DeltaPW}{\Delta\PW}
\newcommand{\alphaQED}{\alpha_\text{QED}}

\newcommand{\syst}{\text{(syst.)}}
\newcommand{\theo}{\text{(theo.)}}
\newcommand{\lumi}{\text{(lumi.)}}

\newcommand{\CMS}{\text{CMS}}
\newcommand{\NP}{\text{NP}}
\newcommand{\SM}{\text{SM}}
\newcommand{\UL}{\text{UL}}

\newcommand{\Ztofl}{{Z \to 4\ell}}
\newcommand{\Ztoll}{Z \to \lplm}
\newcommand{\ppto}{pp \to}
\newcommand{\qqto}{q \bar{q} \to}
\newcommand{\BF}{\sB(\Ztofl)}
\newcommand{\BUL}{\sB_\UL}

\newcommand{\keV}{\kilo\electronvolt}
\newcommand{\GeV}{\giga\electronvolt}
\newcommand{\TeV}{\tera\electronvolt}
\newcommand{\fb}{\femto\barn}
\newcommand{\invfb}{\per\fb}
\newcommand{\pb}{\pico\barn}

\newcommand{\MadGraph}{\textsc{MadGraph5}}
\newcommand{\MadGraphaMC}{\textsc{MadGraph5\_aMC@NLO}}
\newcommand{\FeynRules}{\textsc{FeynRules}}

\newcommand{\ourval}{(4.58 \pm 0.26) \times 10^{-6}}
\newcommand{\uncorr}{(4.59 \pm 0.25) \times 10^{-6}}
\newcommand{\corr}{(4.56 \pm 0.27) \times 10^{-6}}
\newcommand{\uplim}{\num{5.01e-6}}
\newcommand{\smpw}{11.47 \pm 0.07}
\newcommand{\smbf}{(4.70 \pm 0.03) \times 10^{-6}}
\newcommand{\uncQED}{0.6\%}

\begin{document}

\title{Branching fraction for $Z$ decays to four leptons \\
    and constraints on new physics}
\author{J.~Lovelace Rainbolt\thanks{\href{mailto:jlrainbolt@northwestern.edu}{jlrainbolt@northwestern.edu}} 
    \ and Michael Schmitt\thanks{\href{mailto:m-schmitt@northwestern.edu}{m-schmitt@northwestern.edu}} \\
    \emph{Department of Physics and Astronomy, Northwestern University, Evanston, IL 60208, USA}}
\maketitle

\begin{abstract}
The LHC experiments have measured the branching fraction for $Z$ decays to four leptons (electrons or muons).  We have combined these measurements with the result ${\mathcal{B}(Z \to 4\ell)} = (4.58 \pm 0.26) \times 10^{-6}$, allowing a precise comparison to the standard model prediction of $(4.70 \pm 0.03) \times 10^{-6}$.  We use a minimal extension of the standard model to demonstrate that this combined value may be used to set stringent limits on new physics.
\end{abstract}

%
%

\section{Introduction} \label{sec:intro}

Decays that are rare in the standard model~(SM) provide stringent tests of SM predictions and could, in principle, provide indirect or even direct evidence for new physics beyond the standard model~(BSM).  For these reasons, rare decays are deserving of thorough experimental analysis which should be as precise as possible.

\begin{figure} \centering
	\includegraphics{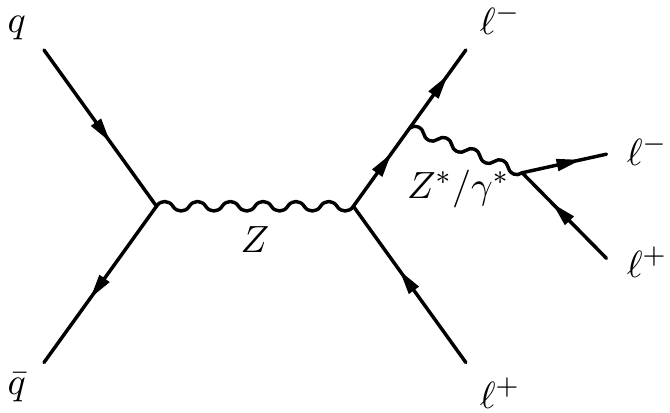}
	\caption{The $\qqto \Ztofl$ process, where $\ell = e, \mu$.}
	\label{SMDiagram}
\end{figure}

An example of a rare SM process is the decay of a single on-shell $Z$ boson to four charged leptons, here denoted $\Ztofl$, where ${\ell = e, \mu}$.  The specific decay we consider, shown in the leading-order Feynman diagram of Fig.~\ref{SMDiagram}, is characterized by the emission of a virtual photon or $Z$~(denoted $Z^*$) which materializes as an additional lepton pair.  The branching fraction for this process, summed over all possible lepton combinations, is predicted to be of order $10^{-6}$.  Even so, the high luminosities recorded at the LHC experiments have enabled them to observe $\Ztofl$ decay events in numbers sufficient to measure $\BF$ precisely.  These measurements have been made by the CMS Collaboration, which has published results from data samples at $\sqrt{s} = 7$ and \SI{13}{\TeV}~\cite{CMS12, CMS16, CMS18}, and by the ATLAS Collaboration, which has performed a combined measurement using data samples at $\sqrt{s} = 7$ and \SI{8}{\TeV}~\cite{ATLAS}.

All four measurements are in agreement with SM predictions and may be combined to reduce the experimental uncertainty, which is statistically limited.  In the case of $\BF$, such a combination is nontrivial; a naive weighted average of the measured values as reported in Refs.~\cite{CMS12, CMS16, CMS18, ATLAS} falsely suggests disagreement among them.  Care must be taken to account for the particular experimental cuts employed in each analysis in order to appropriately incorporate them in the combined value.

This endeavor is worthwhile because the unique topology of the $\Ztofl$ decay makes it sensitive to new physics that couples to leptons.  This type of new physics may be linked to observed BSM phenomena such as the muon~${g - 2}$ anomaly~\cite{GMinus2} and violations of lepton flavor universality in $B$ decays~\cite{BDecays1, BDecays2}.  Indeed, previous studies have used the LHC measurements of $\BF$ to set limits on hypothesized gauge bosons coupling to leptons, such as the $Z'$ in ${L_\mu - L_\tau}$ models intended to explain recent heavy-flavor anomalies~\cite{MuTauLimits1, MuTauLimits2} and in other leptophilic dark matter models~\cite{LeptophilicLimits}.  These constraints are limited, however, by the experimental uncertainties in the measurements from which they are derived, and would benefit from a more precise value.

We report a combined value for $\BF$ determined from the CMS and ATLAS measurements and demonstrate its efficacy in constraining a toy new physics~(NP) model.  We begin with a review of the published data and our combination procedure, which takes into account differing experimental cuts as well as correlated uncertainties among the measurements.  We present our combined $\BF$ result, which is in close agreement with the SM prediction.  Next, we describe our toy NP model and present the resulting constraints, which we discuss in terms of how they demonstrate the power of a precise value for $\BF$ to potentially constrain more developed NP models.  Finally, we conclude with a summary of our findings.

%
%

\section{Measurement combination} \label{sec:meas}

We define the phase space region for our combined $\BF$ by the invariant mass requirements ${80 < \mfl < \SI{100}{\GeV}}$ for the four-lepton sum and ${\mll > \SI{4}{\GeV}}$ for all opposite-sign, same-flavor lepton pairs.  This region is chosen to most closely match those used for the published measurements, and is identical to that of the CMS results published in 2016~\cite{CMS16} and 2018~\cite{CMS18}.  The region used for the 2012 CMS measurement reported in Ref.~\cite{CMS12}, however, is further restricted to ${m_{\ell\ell} > \SI{4}{\GeV}}$ for \emph{all} lepton pairs (regardless of flavor and charge).  This smaller phase space region has a smaller predicted $\BF$.  In order to more accurately include this measurement in our combination, we scaled it using a kinematic acceptance correction (approximately 90\%) calculated from Monte Carlo~(MC) events generated with {\MadGraphaMC}~2.6.1~\cite{MadGraph}.  The measurement reported by ATLAS in Ref.~\cite{ATLAS} corresponds to yet another region, defined by ${\mll > \SI{5}{\GeV}}$, in which $\BF$ is smaller by approximately 25\%.  However, the authors of Ref.~\cite{ATLAS} also provide a value for $\BF$ scaled to ${\mll > \SI{4}{\GeV}}$, which we use here.  This value is listed in Table~\ref{MeasurementTable} with the other measurements.

\begin{table}[t] \centering
	\begin{tabular}{l l p{.65in} p{.65in} p{.65in} l l l l l}
    \toprule
    		&	&	&	&	& \multicolumn{5}{c}{$\BF$ ($\times 10^{-6}$)} \\
	\cmidrule(lr){6-10}
    Collab. & Pub. & $\sqrt{s}$ (\si{\TeV}) & $\sL$ (\si{\invfb}) & Yield & & (stat.) & (syst.) & (theo.) & (lumi.) \\
    \midrule
    CMS	& 2012 & 7 & 5.02 & 28 & 4.6 & $^{+ 1.0}_{- 0.9}$ & $\pm 0.2$ \\ \addlinespace
		& 2016 & 13 & 2.6 & 39 & 4.9 & $^{+ 0.8}_{- 0.7}$ & $^{+ 0.3}_{- 0.2}$ & $^{+ 0.2}_{- 0.1}$ & $\pm 0.1$ \\ \addlinespace
		& 2018 & 13 & 35.9 & 509 & 4.83 & $^{+ 0.23}_{- 0.22}$ & $^{+ 0.32}_{- 0.29}$ & $\pm 0.08$ & $\pm 0.12$ \\ \addlinespace
    ATLAS	& 2014 & 7 (8) & 4.5 (20.3) & 21 (151) & 4.31 & $\pm 0.34$ & $\pm 0.17$ \\
    \bottomrule
\end{tabular}
	\caption{The CMS~\cite{CMS12, CMS16, CMS18} and ATLAS~\cite{ATLAS} measurements of $\BF$ in the phase space region $80 < \mfl < \SI{100}{\GeV}$, $\mll > \SI{4}{\GeV}$.  We scaled the CMS 2012 result to conform to this region.  The data sets used for the CMS 2016 and 2018 measurements do not overlap.}
	\label{MeasurementTable}
\end{table}

Measurement procedures differ across the analyses.  The ATLAS and 2012 CMS analyses determine $\BF$ by comparing the number of observed signal $\Ztofl$ events to the number of ${\Ztoll}$ events in their respective fiducial regions.  By contrast, the 2016 and 2018 CMS analyses use a likelihood fit to measure the ${\ppto \Ztofl}$ cross section, ${\sigma(\ppto \Ztofl)}$, and extract $\BF$ using a theoretical value for ${\sigma(\ppto \Ztoll)}$.  Hence, these two measurements include separate uncertainties on theory and luminosity, which have no bearing on the other measurements.

\begin{table} \centering
	\bigskip
	\begin{tabular}{p{1in} p{0.75in} p{0.75in}}
    \toprule
    	& \multicolumn{2}{c}{Correlation coefficient~($\rho$)} \\
    \cmidrule(lr){2-3}
    Source & CMS & ATLAS \\
    \midrule
    Statistics & 0 & 0 \\
    Experiment & 0.3 & 0 \\
    Theory & 1 & --- \\
    Luminosity & 0.5 & --- \\
    \bottomrule
\end{tabular}
	\caption{The correlation coefficients for each source of uncertainty.}
	\label{CorrelationTable}
\end{table}

For the combination of the four LHC values, we employ the best linear unbiased estimator~(BLUE) method~\cite{Lyons}.  Each source of uncertainty is given a correlation coefficient~$\rho$.  The choices for these coefficients, listed in Table~\ref{CorrelationTable}, are motivated by knowledge of the measurement process, the experiments, and the LHC.  The only nonzero $\rho$ are among the CMS measurements.  The experimental uncertainty coefficient ${\rho^\syst_\CMS = 0.3}$ is derived from the contributions tabulated in Ref.~\cite{CMS16}, and is defined as the correlated fraction of their sum in quadrature~(i.e.~the variance).  The theory coefficient ${\rho^\theo_\CMS = 1}$ is chosen to account for CMS analysis procedures which are standard across the measurements, and the luminosity coefficient ${\rho^\lumi_\CMS = 0.5}$ is chosen as a conservative estimate for correlations among luminosity uncertainty calculations for different sets of data.

The combined value is
\begin{equation}
	\BF = \ourval
	\label{CombinedBF}
\end{equation}
with a reduced~$\chi_\nu^2 = 0.99$~(80\% probability).  The value in Eq.~(\ref{CombinedBF}) may be compared to the value ${\uncorr}$, obtained by setting all ${\rho = 0}$, or to the extreme case $\rho^\syst_\CMS = \rho^\theo_\CMS = \rho^\lumi_\CMS = 1$, which yields~$\corr$.

\begin{figure} \centering
	\includegraphics[scale=1.375]{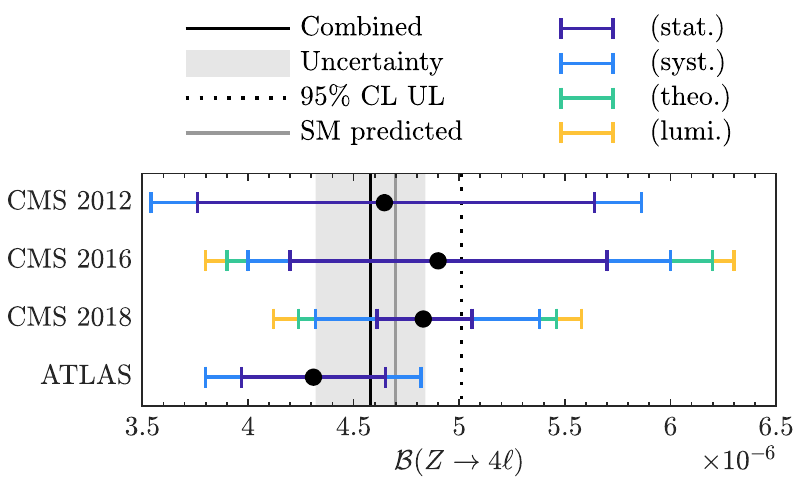}
	\caption{The combined value~(solid black line with gray uncertainty region) compared to the measurements listed in Table~\ref{MeasurementTable}, with colored~(grayscale) error bars corresponding to each source of uncertainty.  Also shown are an upper limit on the combined value at 95\%~CL~(dotted black line) and the SM predicted central value~(solid gray line).  The theoretical uncertainty in the SM value is assessed at $\uncQED$, which is much smaller than the measurement uncertainty.}
	\label{ComparisonPlot}
\end{figure}

The combined $\BF$ value is compared to the LHC measurements in Fig.~\ref{ComparisonPlot}.  The figure also demonstrates the agreement of the combined value with the SM prediction
\begin{equation}
	\frac{\PW}{\Gamma_Z} = \frac{(\smpw)\,\si{\keV}}{\SI{2.441}{\GeV}} = \smbf,
\end{equation}
calculated at leading order~(LO) with {\MadGraph}.  Here, $\PW$ is the partial width for the $\Ztofl$ decay subject to ${\mll > \SI{4}{\GeV}}$, determined from a sample of \num{e5} generated decay events, and $\Gamma_Z$ is the total $Z$ width in the program's default SM implementation, which differs slightly from the measured value.  In order to assess a theoretical uncertainty, we consider the running of $\alphaQED$ as a function of the invariant mass of the less massive opposite-sign, same-flavor lepton pair.  (The running of $\alphaQED$ is not implemented in {\MadGraph}.)  Using the numerical routines of Ref.~\cite{Jegerlehner}, we assess a theoretical systematic uncertainty in $\PW$ of $\uncQED$.

%
%

\section{New physics model}

\begin{figure} \centering
	\includegraphics{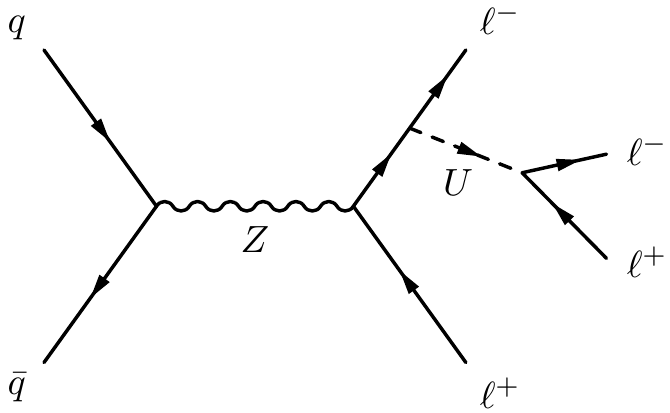}
	\caption{The ${\qqto \Ztofl}$ process mediated by $U$, shown here as a scalar boson.}
	\label{NPDiagram}
\end{figure}

To test the constraining power of the combined $\BF$ value, we developed a ``toy'' NP model.  This is a naive, minimal extension of the SM which includes additional contributions to $\Ztofl$ decay.  These contributions arise from the addition of a single new boson resonance, named $U$, which may appear in the ${\qqto \Ztofl}$ process as shown in Fig.~\ref{NPDiagram}.  The appearance of $U$ in such a decay is, fundamentally, permitted only in the case that $U$ is uncharged, less massive than the $Z$ boson, and coupled to one or both lepton flavors.  These are the only restrictions we consider for the demonstrative purposes of this study.

We have modeled $U$ as both a scalar and a vector boson.  The scalar and vector models, respectively, are described by the additions to the SM Lagrangian
\begin{equation}
    \sL \supset \frac12 \partial_\mu U \partial^\mu U - \frac12 M_U^2 U^2 + g_\ell \bar{\ell} \ell U
    \label{ScalarLagrangian}
\end{equation}
and
\begin{equation}
    \sL \supset \frac12 \partial_\mu U_\nu \partial^\mu U^\nu + \frac12 M_U^2 U_\mu U^\mu + g_\ell \bar{\ell} \gamma_\mu \ell U^\mu,
    \label{VectorLagrangian}
\end{equation}
where $M_U$ is the mass of the $U$ boson and $g_\ell$~(${= g_e, g_\mu}$) the strengths of its lepton couplings.  We implement Eqs.~(\ref{ScalarLagrangian}, \ref{VectorLagrangian}) with {\FeynRules}~2.3.29~\cite{FeynRules}, allowing $M_U$ and $g_\ell$ to vary over the parameter space defined by ${1 \leq M_U \leq \SI{90}{\GeV}}$ and ${0.005 \leq g_\ell \leq 0.45}$.  The full width of $U$,~$\Gamma_U$, depends on these parameters and is calculated automatically by {\MadGraph}, which we use to generate MC events at LO.  For the scalar~(vector) model, ${\num{1.9e-6} \leq \Gamma_U \leq \SI{1.5}{\GeV}}$ ($\num{5.3e-6} \leq \Gamma_U \leq \SI{3.9}{\GeV}$).  The largest values of $\Gamma_U$ in these ranges are eventually excluded.  For all values considered here, the leptons are prompt.

To explore the models' behavior as a function of the input parameters, we examined the quantity~$\DeltaPW$, defined as the deviation of the $\Ztofl$ partial width from the SM value.  That is,
\begin{equation}
    \DeltaPW \equiv \PW^{\NP} - \PW^{\SM},
    \label{PartialWidthEq}
\end{equation}
where the quantity denoted ``\NP'' is calculated from the NP model for a given choice of free parameters, and that denoted ``\SM'' is the SM prediction.  These calculations were performed with {\MadGraph}, which takes quantum mechanical interference into account.  $\DeltaPW$ is almost always positive.

\begin{figure} \centering
	\begin{tabular}{l}
	\includegraphics[scale=1.375]{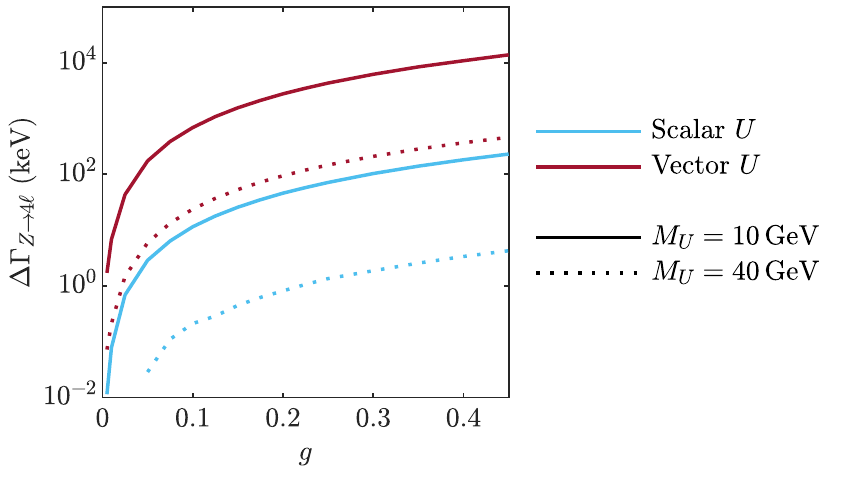} \\
	\\
	\includegraphics[scale=1.375]{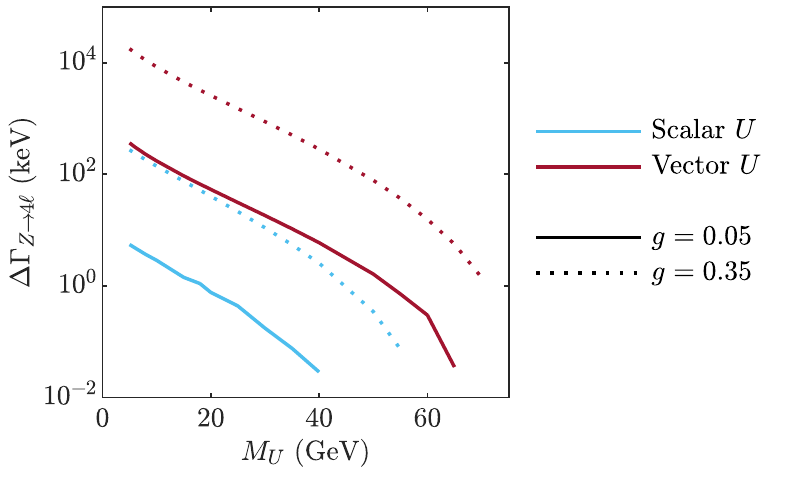} \\
	\end{tabular}
	\caption{The dependence of $\DeltaPW$ on $g$~(top) and $M_U$~(bottom) for representative values of the other parameter, one ``realistic''~(solid line) and one ``extreme''~(dotted line).  The scalar $U$ model is shown in blue~(light gray) and the vector $U$ model in red~(dark gray).}
	\label{ParameterPlots}
\end{figure}

Examples of the behavior of $\DeltaPW$ as a function of $M_U$ and $g$~(${\equiv g_e = g_\mu}$) are shown in Fig.~\ref{ParameterPlots}.  For both models, $\DeltaPW$ exhibits the expected power-law dependence on each parameter.

%
%

\section{Constraints}

In order to apply constraints from the measurements to the NP models, we define an acceptance for the fiducial region used for the 2016 and 2018 CMS measurements~\cite{CMS16, CMS18}, which correspond to our chosen phase space.  In addition to the phase space requirements ${80 < \mfl < \SI{100}{\GeV}}$ and ${\mll > \SI{4}{\GeV}}$, this fiducial region is defined by the transverse momentum requirements ${p_T^{\ell_1} > \SI{20}{\GeV}}, \ {p_T^{\ell_2} > \SI{10}{\GeV}}$, and ${p_T^{\ell_{3,4}} > \SI{5}{\GeV}}$; the pseudorapidity requirement ${|\eta^\ell| < 2.5}$ for each lepton; and the invariant mass requirement ${m_{Z_1} > \SI{40}{\GeV}}$, where $Z_1$ is the more massive opposite-sign, same-flavor lepton pair.  We define our exclusion as the parameter space region such that
\begin{equation}
    \sigma_\NP A_\NP > \frac{\BUL}{\sB} \sigma_\SM A_\SM,
\end{equation}
where ${\sigma = \sigma(\ppto \Ztofl)}$ is the cross section and $A$~the acceptance, and the labels~``\NP'' and~``\SM'' are used as in Eq.~\ref{PartialWidthEq}.  The quantity~$\sB$ is our combined branching fraction given in Eq.~\ref{CombinedBF}, and $\BUL$ is a corresponding upper limit.  At the 95\% confidence level~(CL), for instance, $\BUL = \uplim$.

\begin{figure} \centering
	\includegraphics[scale=1.375]{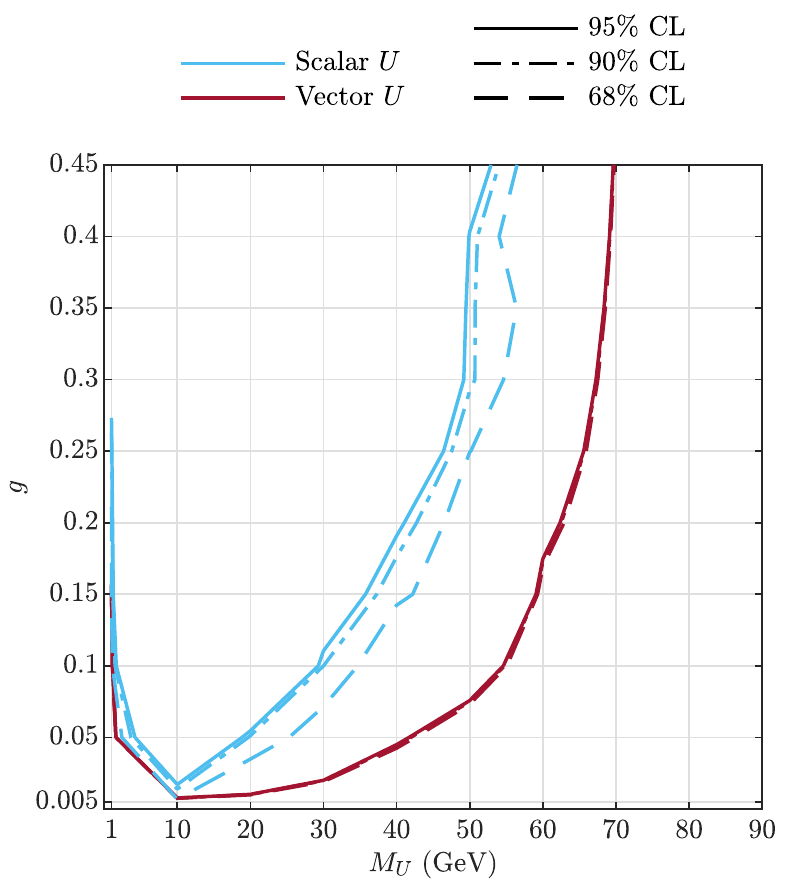}
	\caption{Excluded regions in ${(M_U, g)}$ space when ${g_e = g_\mu}$~($\equiv g$), compared at 95\%, 90\%, and 68\%~CLs.  The scalar $U$ model is shown in blue~(light gray) and the vector $U$ model in red~(dark gray).}
	\label{ExclusionCurves}
\end{figure}

\begin{figure} \centering
	\includegraphics[scale=1.375]{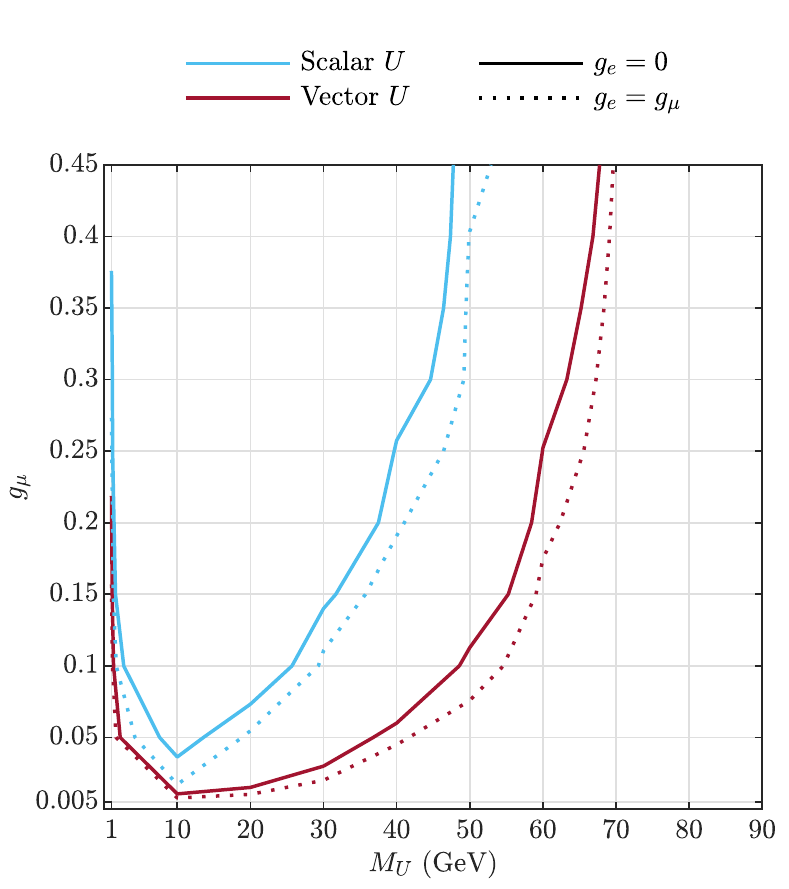}
	\caption{Excluded regions in $(M_U, g_\mu)$ space for the cases $g_e = 0$~(solid line) and $g_e = g_\mu$~(dotted line), compared at 95\%~CL.  The scalar $U$ model is shown in blue~(light gray) and the vector $U$ model in red~(dark gray).}
	\label{MuonOnlyExclusionCurves}
\end{figure}

Exclusion curves were interpolated from a parameter space sampled in intervals of \SI{10}{\GeV} in $M_U$ and 0.05 in $g$.  Approximately \num{2e4} ${\ppto \Ztofl}$ MC events were generated at each sampled point, and the quantities $\sigma_\NP$ and $A_\NP$ calculated from them at the generator level.  For the scalar~(vector) model, they fall in the ranges ${0.186 \leq \sigma_\NP \leq \SI{3.66}{\pb}}$ (${\num{1.09e-9} \leq \sigma_\NP \leq \SI{209}{\pb}}$) and ${9.9 \leq A_\NP \leq 14.3\%}$ (${3.7 \leq A_\NP \leq 32.6\%}$).  The SM quantities ${\sigma_\SM = \SI{194}{\fb}}$ and ${A_\SM = 10.5\%}$ were obtained by the same procedure.  Figure~\ref{ExclusionCurves} shows the exclusion curves for both models in the case ${g_e = g_\mu}$~($\equiv g$), corresponding to $\BUL$ at 95\%, 90\%, and 68\%~CLs.  Figure~\ref{MuonOnlyExclusionCurves} compares the $g_e = g_\mu$ case to the case ${g_e = 0}$ at 95\%~CL.

%
%

\section{Discussion}

As illustrated in Fig.~\ref{ComparisonPlot}, our combined value for $\BF$ lies well within the range of the measurements listed in Table~\ref{MeasurementTable}.  The \emph{total} uncertainty in the combined value is comparable to the \emph{statistical} uncertainty in the most recent result published by CMS~\cite{CMS18}, and is smaller than the statistical uncertainties in all of the other measurements.  This reduced uncertainty leads to a tight upper limit on $\BF$ and the exclusion of a sizable region in the parameter space of our toy NP models.  The boundary of this excluded region does not depend strongly on the CL of the upper limit, especially for the vector $U$ model, as illustrated in Fig.~\ref{ExclusionCurves}.  Although these toy models are incomplete by construction, they are suggestive of how the combined value for $\BF$ could be used to constrain more sophisticated models like those mentioned in Sec.~\ref{sec:intro}.

However, holistic statements about the constraining power of $\BF$ as it would be applied to any developed model must consider experimental results from elsewhere, such as studies done at LEP.  A complete discussion of these results is largely model dependent and therefore beyond the scope of this study.  As an illustration, however, we may use our toy NP models to consider the case where $U$ does not couple to electrons at all.  LEP results provide little or no evidence for a $U$ boson that couples to electrons~\cite{LEPLimits}.  While $\BF$ includes contributions from the nonconstructive ${Z \to 4e}$ and ${Z \to 2e2\mu}$ channels in this case, it still offers a valuable test of the SM in the ${Z \to 4\mu}$ channel.  As shown in Fig.~\ref{MuonOnlyExclusionCurves}, imposing the restriction ${g_e = 0}$ shrinks the excluded region only slightly.

Constraints from $\BF$ should be applied to future NP searches at the LHC.  These searches would complement those already performed at lower-energy experiments, such as \emph{BABAR}; the collaboration has published limits on light $Z'$ bosons with $g_\mu$ as low as \num{7e-4}~\cite{BABAR}, corresponding to ${M_U < \SI{10}{\GeV}}$ in our parameter space.  Similarly, the KLOE-2 Collaboration has set stringent limits on dark photons of masses below ${\SI{1}{\GeV}}$~\cite{KLOE2}.

%
%

\section{Conclusion}

We have combined the CMS~\cite{CMS12, CMS16, CMS18} and ATLAS~\cite{ATLAS} measurements of the $Z$ branching fraction to four leptons, $\BF$ where ${\ell = e, \mu}$, with the result ${\BF = \ourval}$.  This combined value corresponds to the phase space ${80 < \mfl < \SI{100}{\GeV}}$, ${\mll > \SI{4}{\GeV}}$ and takes into account nonuniformity in the experimental cuts employed by the analyses, as well as correlated systematic uncertainties among the measurements.  This combined value agrees well with the SM prediction ${\BF = \smbf}$.

We have used this combined value to constrain the parameters of a toy NP model including a new scalar or vector boson~$U$, which may couple to electrons and muons.  These constraints exclude a large region of the parameter space~${(M_U, g_\ell)}$, where $M_U$ is the mass of the $U$ boson and $g_\ell$ its lepton couplings.  These limits demonstrate the potential of the combined $\BF$ value to constrain theoretically- and experimentally-informed NP models, as well as searches for BSM physics at the LHC.

\paragraph*{Note added.}
After this work was completed, the CMS Collaboration released a search for a $Z'$ boson in the $Z \to 4\mu$ channel~\cite{EXO}.  The search targets a $Z'$ boson that is described by ${L_\mu - L_\tau}$ models and has a mass between 5 and \SI{70}{\GeV}.  The authors set limits on the boson's mass and coupling constant, and compare those limits to constraints from $\BF$ quoted in Ref.~\cite{MuTauLimits1}.

%
%

\section*{Acknowledgments}

We would like to thank I.~Low for theoretical discussions.  We would also like to thank S.~Bhattacharya and B.~Pollack for useful comments on the manuscript.  We gratefully acknowledge the support provided by the Department of Energy under award number DE-SC0015910.

%
%

\providecommand{\href}[2]{#2}\begingroup\raggedright\endgroup


\begin{thebibliography}{10}%
\makeatletter
\providecommand{\hrefCMSnoop }[0]{\@secondoftwo}%
\makeatother
\providecommand{\doi}{\texttt{doi:}\begingroup \urlstyle{tt}\Url}

\bibitem{CMS12}
\hrefCMSnoop {}{{CMS Collaboration}, ``Observation of {$Z$} decays to four
  leptons with the {CMS} detector at the {LHC}'',} \textit{ JHEP} \textbf{ 12}
  (2012) 034,
  \href{http://dx.doi.org/10.1007/JHEP12(2012)034}{\doi{10.1007/JHEP12(2012)034}},
\href{http://www.arXiv.org/abs/1210.3844}{\texttt{arXiv:1210.3844}}.

\bibitem{CMS16}
\hrefCMSnoop {}{{CMS Collaboration}, ``Measurement of the {$ZZ$} production
  cross section and {$Z \to \ell^+ \ell^- \ell'^+ \ell'^-$} branching fraction
  in {$pp$} collisions at {$\sqrt s = 13$\,~TeV}'',} \textit{ Phys. Lett.}
  \textbf{ B763} (2016) 280--303,
  \href{http://dx.doi.org/10.1016/j.physletb.2016.10.054}{\doi{10.1016/j.physletb.2016.10.054}},
  \href{http://www.arXiv.org/abs/1607.08834}{\texttt{arXiv:1607.08834}}.
Erratum: \textit{Phys. Lett.}, B772:884, 2017.

\bibitem{CMS18}
\hrefCMSnoop {}{{CMS Collaboration}, ``Measurements of the {$pp \to ZZ$}
  production cross section and the {$Z \to 4\ell$} branching fraction, and
  constraints on anomalous triple gauge couplings at {$\sqrt{s} =
  13$\,~TeV}'',} \textit{ Eur. Phys. J.} \textbf{ C78} (2018) 165,
  \href{http://dx.doi.org/10.1140/epjc/s10052-018-5567-9}{\doi{10.1140/epjc/s10052-018-5567-9}},
\href{http://www.arXiv.org/abs/1709.08601}{\texttt{arXiv:1709.08601}}.

\bibitem{ATLAS}
\hrefCMSnoop {}{{ATLAS Collaboration}, ``Measurements of four-lepton production
  at the {$Z$} resonance in {$pp$} collisions at {$\sqrt{s} = 7$} and {8\,~TeV}
  with {ATLAS}'',} \textit{ Phys. Rev. Lett.} \textbf{ 112} (2014), no.~23,
  231806,
  \href{http://dx.doi.org/10.1103/PhysRevLett.112.231806}{\doi{10.1103/PhysRevLett.112.231806}},
\href{http://www.arXiv.org/abs/1403.5657}{\texttt{arXiv:1403.5657}}.

\bibitem{GMinus2}
\hrefCMSnoop {}{J.~P. Miller, E.~de~Rafael, B.~L. Roberts, and
  D.~{St\"{o}ckinger}, ``Muon {$(g - 2)$}: Experiment and theory'',} \textit{
  Ann. Rev. Nucl. Part. Sci.} \textbf{ 62} (2012) 237--264,
\href{http://dx.doi.org/10.1146/annurev-nucl-031312-120340}{\doi{10.1146/annurev-nucl-031312-120340}}.

\bibitem{BDecays1}
\hrefCMSnoop {}{S.~Bifani, S.~Descotes-Genon, A.~Romero~Vidal, and M.-H.
  Schune, ``Review of lepton universality tests in {$B$} decays'',} \textit{ J.
  Phys.} \textbf{ G46} (2019), no.~2, 023001,
  \href{http://dx.doi.org/10.1088/1361-6471/aaf5de}{\doi{10.1088/1361-6471/aaf5de}},
\href{http://www.arXiv.org/abs/1809.06229}{\texttt{arXiv:1809.06229}}.

\bibitem{BDecays2}
\hrefCMSnoop {}{{HFLAV Collaboration}, ``Averages of {$b$}-hadron,
  {$c$}-hadron, and {$\tau$}-lepton properties as of summer 2016'',} \textit{
  Eur. Phys. J.} \textbf{ C77} (2017), no.~12, 895,
  \href{http://dx.doi.org/10.1140/epjc/s10052-017-5058-4}{\doi{10.1140/epjc/s10052-017-5058-4}},
\href{http://www.arXiv.org/abs/1612.07233}{\texttt{arXiv:1612.07233}}.

\bibitem{MuTauLimits1}
\hrefCMSnoop {}{W.~Altmannshofer, S.~Gori, S.~Profumo, and F.~S. Queiroz,
  ``Explaining dark matter and {$B$} decay anomalies with an {$L_\mu - L_\tau$}
  model'',} \textit{ JHEP} \textbf{ 12} (2016) 106,
  \href{http://dx.doi.org/10.1007/JHEP12(2016)106}{\doi{10.1007/JHEP12(2016)106}},
\href{http://www.arXiv.org/abs/1609.04026}{\texttt{arXiv:1609.04026}}.

\bibitem{MuTauLimits2}
\hrefCMSnoop {}{W.~Altmannshofer, S.~Gori, M.~Pospelov, and I.~Yavin, ``Quark
  flavor transitions in {$L_\mu - L_\tau$} models'',} \textit{ Phys. Rev.}
  \textbf{ D89} (2014) 095033,
  \href{http://dx.doi.org/10.1103/PhysRevD.89.095033}{\doi{10.1103/PhysRevD.89.095033}},
\href{http://www.arXiv.org/abs/1403.1269}{\texttt{arXiv:1403.1269}}.

\bibitem{LeptophilicLimits}
\hrefCMSnoop {}{N.~F. Bell, Y.~Cai, R.~K. Leane, and A.~D. Medina,
  ``Leptophilic dark matter with {$Z'$} interactions'',} \textit{ Phys. Rev.}
  \textbf{ D90} (2014), no.~3, 035027,
  \href{http://dx.doi.org/10.1103/PhysRevD.90.035027}{\doi{10.1103/PhysRevD.90.035027}},
\href{http://www.arXiv.org/abs/1407.3001}{\texttt{arXiv:1407.3001}}.

\bibitem{MadGraph}
J.~Alwall\hrefCMSnoop {}{ {et~al.}, ``The automated computation of tree-level
  and next-to-leading order differential cross sections, and their matching to
  parton shower simulations'',} \textit{ JHEP} \textbf{ 07} (2014) 079,
  \href{http://dx.doi.org/10.1007/JHEP07(2014)079}{\doi{10.1007/JHEP07(2014)079}},
\href{http://www.arXiv.org/abs/1405.0301}{\texttt{arXiv:1405.0301}}.

\bibitem{Lyons}
\hrefCMSnoop {}{L.~Lyons, D.~Gibaut, and P.~Clifford, ``How to combine
  correlated estimates of a single physical quantity'',} \textit{ Nucl.
  Instrum. Meth.} \textbf{ A270} (1988) 110,
\href{http://dx.doi.org/10.1016/0168-9002(88)90018-6}{\doi{10.1016/0168-9002(88)90018-6}}.

\bibitem{Jegerlehner}
\hrefCMSnoop {}{F.~Jegerlehner, ``Precision measurements of
  {$\sigma_\text{hadronic}$} for {$\alpha_\text{eff}(E)$} at {ILC} energies and
  {$(g-2)_\mu$}'',} \textit{ Nucl. Phys. (Proc. Suppl.)} \textbf{ B162} (2006)
  22--32,
  \href{http://dx.doi.org/10.1016/j.nuclphysbps.2006.09.060}{\doi{10.1016/j.nuclphysbps.2006.09.060}},
\href{http://www.arXiv.org/abs/0608329}{\texttt{arXiv:0608329}}.

\bibitem{FeynRules}
A.~Alloul\hrefCMSnoop {}{ {et~al.}, ``{\textsc{FeynRules} 2.0---A} complete
  toolbox for tree-level phenomenology'',} \textit{ Comput. Phys. Commun.}
  \textbf{ 185} (2014) 2250--2300,
  \href{http://dx.doi.org/10.1016/j.cpc.2014.04.012}{\doi{10.1016/j.cpc.2014.04.012}},
\href{http://www.arXiv.org/abs/1310.1921}{\texttt{arXiv:1310.1921}}.

\bibitem{LEPLimits}
\hrefCMSnoop {}{F.~del Aguila, M.~Chala, J.~Santiago, and Y.~Yamamoto,
  ``Collider limits on leptophilic interactions'',} \textit{ JHEP} \textbf{ 03}
  (2015) 059,
  \href{http://dx.doi.org/10.1007/JHEP03(2015)059}{\doi{10.1007/JHEP03(2015)059}},
\href{http://www.arXiv.org/abs/1411.7394}{\texttt{arXiv:1411.7394}}.

\bibitem{BABAR}
\hrefCMSnoop {}{{\emph{BABAR} Collaboration}, ``Search for a muonic dark force
  at \emph{{BABAR}}'',} \textit{ Phys. Rev.} \textbf{ D94} (2016), no.~1,
  011102,
  \href{http://dx.doi.org/10.1103/PhysRevD.94.011102}{\doi{10.1103/PhysRevD.94.011102}},
\href{http://www.arXiv.org/abs/1606.03501}{\texttt{arXiv:1606.03501}}.

\bibitem{KLOE2}
\hrefCMSnoop {}{{KLOE-2 Collaboration}, ``Combined limit on the production of a
  light gauge boson decaying into {$\mu^+ \mu^-$} and {$\pi^+ \pi^-$}'',}
  \textit{ Phys. Lett.} \textbf{ B784} (2018) 336--341,
  \href{http://dx.doi.org/10.1016/j.physletb.2018.08.012}{\doi{10.1016/j.physletb.2018.08.012}},
\href{http://www.arXiv.org/abs/1807.02691}{\texttt{arXiv:1807.02691}}.

\bibitem{EXO}
\hrefCMSnoop {}{{CMS Collaboration}, ``Search for an {$L_\mu - L_\tau$} gauge
  boson using {$Z \to 4\mu$} events in proton-proton collisions at {$\sqrt s =
  13$\,~TeV}'',} (2018).
  \href{http://www.arXiv.org/abs/1808.03684}{\texttt{arXiv:1808.03684}}.
Submitted to Phys. Lett. {\textbf{B}}.

\end{thebibliography}
\end{document}